\documentclass[pra,twocolumn,showpacs,preprintnumbers,amsmath,amssymb,aps]{revtex4}
\usepackage{color}
\usepackage{bm, graphicx, amsmath}

\usepackage{braket}

\begin{document}
	\title{Discrete outcome quantum sensor networks}
	\author{Mark Hillery$^{1,2}$, Himanshu Gupta$^{3}$, Caitao Zhan$^{3}$}
	\affiliation{$^{1}$Department of Physics and Astronomy, Hunter College of the City University of New York, 695 Park Avenue, New York, NY 10065 \\
		$^{2}$Graduate Center of the City University of New York, 365 Fifth Avenue, New York, NY 10016,\\
		$^{3}$ Department of Computer Science, Stony Brook University, Stony Brook, NY 11794}

	\begin{abstract}
		We model a quantum sensor network using techniques from quantum state discrimination.  The interaction between a qubit detector and the environment is described by a unitary operator, and we will assume that at most one detector does interact.  The task is to determine which one does or if none do.    This involves choosing an initial state of the detectors and a measurement.  We consider global measurements in which all detectors are measured simultaneously.  We find that an entangled initial state can improve the detection probability, but this advantage decreases as the number of detectors increases.
	\end{abstract}
	
	\maketitle
	
	\section{Introduction}
	
	The question of whether quantum mechanics can improve the behavior of sensor networks is one that is attracting considerable attention.  It is known that it can be used to improve the performance of individual detectors, perhaps the most spectacular example being the use of squeezed states to improve the sensitivity of the LIGO gravitational wave detector.  When there is a network of detectors, new questions arise, in particular, whether entanglement between the detectors can enhance the performance of the network.  The answers so far are mixed. 
	
	Previous work on quantum sensor networks has studied the following problem.  The detectors, which are quantum systems, have, as a result of their interaction with the environment, parameters, encoded in their state.  In all cases, these parameters have been taken to be continuous variables.  For example, these parameters could be the strength of a magnetic field at different locations.  We would like to estimate these parameters or some function of them.  The detectors can be qubits \cite{gorshkov,qian}, continuous variable systems \cite{shapiro}, or general quantum systems \cite{proctor}.  It has been found that for finite-dimensional systems, entanglement of the quantum systems does not provide an advantage in estimating the individual parameters, but does provide an advantage in estimating a function of them \cite{gorshkov,qian,proctor,Rubio}.  It has been shown that entangled states in optical networks can provide an advantage for distributed sensing \cite{shapiro,Zhang}.  Further studies have investigated whether linear optical networks with unentangled inputs can give a quantum advantage in distributed metrology \cite{ge}, and whether continuous-variable error correction can be useful in protecting a network of continuous-variable sensors from the effects of noise \cite{preskill}.
	
	Suppose that instead of determining a parameter, one is interested in whether a detector has detected something or which detector has detected something.  This kind of problem is described by discrete rather than continuous variables, and is a problem in channel discrimination \cite{kitaev,acin,DAriano,sacchi1,sacchi2,wang,pirandola}.  Each detector in a network can receive an input or no input.  For example, the detectors could be designed to detect magnetic fields, and if a field is present at the location of a detector, its state would be altered.  Another possibility is that weak coherent states or light could be sent out and reflected back, and if a transparent medium were present in the path, the coherent state received back would be different than if no medium were present.  Now suppose the unitary operator $U$ describes the interaction between an input and a detector, and that only one detector in the network has received an input, but we do not know which one.  For example, $U$ could describe the rotation of a spin caused by a magnetic field, or a phase shift induced in a state of light by a transparent object.  The different output states of the detector will be produced by $U$ from one of the detectors and the identity from the rest acting on the initial state of the detectors.  We then want to measure the output state in order to determine which detector received an input.  This means that we have to optimize over both the initial state of the detectors, and the final measurement.  The related problem of picking out a target quantum channel from a background of identical channels has been analyzed by Zhuang and Pirandola, and useful bounds on channel discrimination have been derived \cite{Zhuang1,Zhuang2,Pereira}.
	
	We will begin by analyzing a two-detector network.  We will look at two measurement schemes for the final state of the detectors, minimum  error and unambiguous.  Minimum error discrimination always returns an answer, but the answer can be wrong.  The probability of making a mistake is, however, minimized.  In unambiguous discrimination, there are no errors, but measurements can fail.  Which scheme is used depends on the relative cost of making a mistake versus receiving no answer.  The case in which one detector has interacted and the measurement is of the minimum-error type can be solved by using a method developed in \cite{DAriano}.  We will also examine the case in which either no detectors have interacted or only one has, and our task is to decide which of these two alternatives has occurred.  We next look at $N$ detectors.  
	
	\section{Two detectors}
	\subsection{Finding which detector interacted}
	We have 2 detectors, each of which is a qubit, initially in the two-qubit state $|\psi\rangle$.  One of the detectors has its state altered by an interaction with the environment, and we would like to know which.  That means we wish to discriminate between the states $|\phi_{1}\rangle = (U\otimes I)|\psi\rangle$ and $|\phi_{2}\rangle = (I\otimes U)|\psi\rangle$, where the unitary operator, $U$, describes the interaction.    This leads to two questions.  How should we choose $|\psi\rangle$ and how should we choose a measurement to accomplish this optimally?  This leads us to a channel discrimination problem, where we wish to discriminate between $U\otimes I$ and $I\otimes U$.
	
	There are two possible ways to approach this problem.  In general, the states  $(U\otimes I)|\psi\rangle$ and $(I\otimes U)|\psi\rangle$ will not be orthogonal, which means they cannot be distinguished perfectly.  One possibility, minimum-error discrimination, is to have the possibility of making a mistake, but to minimize the probability that this occurs.  The probability of succeeding in that case is \cite{helstrom}
	\begin{equation} 
	P_{s}^{(min)} = \frac{1}{2} \left( 1 + \sqrt{ 1 - |\langle \psi |(U \otimes U^{-1}) |\psi\rangle |^{2}} \right) .
	\end{equation}
	Another, unambiguous discrimination, is to never make a  mistake, but to allow the measurement to sometimes fail, that is, give no answer.  In this case, one minimizes the failure probability.  The optimal state $|\psi\rangle$ for both strategies is the one that minimizes $|\langle \psi |(U \otimes U^{-1}) |\psi\rangle |$ \cite{review}.
	
	Our two-detector problem is easy to solve, but its solution illustrates some of the features we expect to see in more elaborate situations  \cite{DAriano}.  Without loss of generality we can suppose that the eigenvalues and eigenvectors of $U$ are $U|u_{\pm}\rangle = e^{\pm i\theta} |u_{\pm}\rangle$.  The eigenvalues of $U\otimes U^{-1}$ are then $e^{2i\theta}$, $e^{-2i\theta}$, and $1$ (twice), and throughout this paper we will take $\theta$ to be in the range $-\pi /4\leq \theta \leq \pi /4$.  We can express $|\psi\rangle$ as
	\begin{eqnarray}
	|\psi\rangle & = & c_{++}|u_{+}\rangle |u_{+}\rangle + c_{+-} |u_{+}\rangle |u_{-}\rangle + c_{-+}|u_{-}\rangle |u_{+}\rangle  \nonumber \\
	&&+ c_{--} |u_{-}\rangle |u_{-}\rangle ,
	\end{eqnarray}
	and let
	\begin{eqnarray}
	z & = &  \langle \psi |(U \otimes U^{-1}) |\psi\rangle  \nonumber \\
	& = & |c_{+-}|^{2} e^{2i\theta} + (|c_{++}|^{2} + |c_{--}|^{2}) + |c_{-+}|^{2} e^{-2i\theta} .
	\end{eqnarray}
	The possible values of $z$ lie in the triangle in $\mathbb{C}$ whose vertices are $e^{\pm2i\theta}$ and $1$.  The states that minimize $|\langle \psi |(U \otimes U^{-1}) |\psi\rangle |$ correspond to the value of $z$ that is closest to the origin \cite{DAriano}.   For $0 \leq \theta \leq \pi /4$, this implies that $z$ is on the line connecting $e^{2i\theta}$ and $e^{-2i\theta}$ on the real axis.  Therefore, $c_{++}=c_{--}=0$, and $|c_{+-}|=|c_{-+}|$, so we can choose as the optimal $|\psi\rangle$, for both strategies, the entangled state
	\begin{equation}
	\label{psi-2det}
	|\psi\rangle= \frac{1}{\sqrt{2}} (|u_{+}\rangle |u_{-}\rangle + |u_{-}\rangle |u_{+}\rangle ).
	\end{equation}
	The optimal measurements are global ones, that is, both qubits are measured together, not individually, and the operators describing the measurements are proportional to projections onto entangled states.  This is true for both the minimum-error and unambiguous strategies, though the measurement operators are different in the two cases.  In the minimum-error case, the measurement operators are orthogonal projections and are given by \cite{review}
	\begin{eqnarray} 
	\label{2state-minerr}
	\Pi_{1} & = & |v_{1}\rangle \langle v_{1}| \nonumber \\
	\Pi_{2} & = & |v_{2}\rangle \langle v_{2}| .
	\end{eqnarray}
	where
	\begin{eqnarray}
	|v_{1}\rangle & = & \frac{1}{\sqrt{2}} (|u_{+}\rangle |u_{-}\rangle -i |u_{-}\rangle |u_{+}\rangle ) \nonumber \\
	|v_{2}\rangle & = & \frac{1}{\sqrt{2}} (|u_{+}\rangle |u_{-}\rangle + i |u_{-}\rangle |u_{+}\rangle ) .
	\end{eqnarray}
	The success probability, that is, the probability of getting the right answer, is
	\begin{equation}
	P_{s}^{(min)} = \sum_{j=1}^{2} \frac{1}{2} \langle\phi_{j}|\Pi_{j}|\phi_{j}\rangle = \frac{1}{2}[ 1 + \sin (2\theta )] ,
	\end{equation}
	where we have assumed that the two states are equally likely.  Note that in this case the measurement operators project onto entangled states and are independent of $\theta$.  This latter property means that the same initial state and measurement can be used to determine which detector has interacted with the environment for a range of interaction strengths or times.  
	
	For unambiguous discrimination we are also discriminating between the states $|\phi_{1}\rangle = (U\otimes I)|\psi\rangle$ and $|\phi_{2}\rangle = (I\otimes U)|\psi\rangle$, with $|\psi\rangle$ given by Eq.\ (\ref{psi-2det}).  In order to construct the measurement operators for unambiguous discrimination in the case  $0\leq \theta \leq \pi /4$, we need to define the states
	\begin{eqnarray}
	|\phi_{1}^{\perp}\rangle & = & \frac{1}{\sqrt{2}} ( e^{i\theta} |u_{+}\rangle |u_{-}\rangle - e^{-i\theta} |u_{-}\rangle |u_{+}\rangle  ) \nonumber \\
	|\phi_{2}^{\perp} \rangle& = & \frac{1}{\sqrt{2}} ( e^{-i\theta} |u_{+}\rangle |u_{-}\rangle - e^{i\theta} |u_{-}\rangle |u_{+}\rangle )  .
	\end{eqnarray}
	Note that $\langle \phi^{\perp}_{1}|\phi_{1}\rangle =0$ and $\langle \phi_{2}^{\perp}|\phi_{2}\rangle = 0$.  The operator $\Pi_{1}$ which corresponds to detecting $|\phi_{1}\rangle$, is $\Pi_{1}= d |\phi_{2}^{\perp}\rangle\langle\phi_{2}^{\perp}|$, for some constant $d$, and $\Pi_{2}$ which corresponds to detecting $|\phi_{2}\rangle$, is $\Pi_{2}= d |\phi_{1}^{\perp}\rangle\langle\phi_{1}^{\perp}|$, where we are assuming the states have the same failure probability.  The operator corresponding to the measurement failing is given by
	\begin{equation}
	\Pi_{f}=I - \Pi_{1} - \Pi_{2} ,
	\end{equation}
	and the constant $d$ is determined by the requirement that it be the largest value for which this operator is positive.  The operator $|\phi_{1}^{\perp}\rangle\langle\phi_{1}^{\perp}|+ |\phi_{2}^{\perp}\rangle\langle\phi_{2}^{\perp}|$ has eigenvalues $\lambda = 1 \pm \cos (2\theta )$, which implies that
	\begin{equation}
	d = \frac{1}{1+ \cos (2\theta )} .
	\end{equation}
	Consequently,  we have
	\begin{eqnarray}
	\Pi_{1} & = &\frac{1}{1+ \cos (2\theta )} |\phi_{2}^{\perp}\rangle\langle\phi_{2}^{\perp}|   \nonumber \\
	\Pi_{2} & = & \frac{1}{1+ \cos (2\theta )} |\phi_{1}^{\perp}\rangle\langle\phi_{1}^{\perp}|  . 
	\end{eqnarray}
	The probability of the measurement succeeding is
	\begin{equation}
	P_{s}^{(un)} = 1-|\cos (2\theta )| .
	\end{equation} 
	
	\subsection{One or none}
	The measurements made and the initial state of the detectors depends on the question asked.  As an illustration, let us continue to look at two detectors but ask a different question.  Suppose we are only interested in whether one of the detectors has fired or neither has.  We will assume that the probability of neither firing is $p_{0}$, that each detector has an equal probability of firing, that the probability of both firing is sufficiently small that it can be neglected, and that the probability that one or the other detector fires is $p_{1}=1-p_{0}$.  This can be summarized by saying that we want to discriminate between the density matrices 
	\begin{eqnarray}
	\rho_{0} & = & |\psi\rangle\langle\psi | \nonumber \\
	\rho_{1} & = & \frac{1}{2} [ (U\otimes I)|\psi\rangle\langle\psi |(U^{\dagger}\otimes I) \nonumber \\
	&& + (I\otimes U)|\psi\rangle\langle\psi |(I\otimes U^{\dagger})  ] ,
	\end{eqnarray}
	where $|\psi\rangle$ is the initial state of the detectors, and $\rho_{0}$ occurs with probability $p_{0}$ and $\rho_{1}$ occurs with probability $p_{1}$.  The optimal success probability for the minimum error measurement in this more general case is given by
	\begin{equation}
	P_{s}^{(min)} = \frac{1}{2} ( 1 + \|\Lambda\|_{1}) 
	\end{equation}
	where $\Lambda = p_{0}\rho_{0} - p_{1}\rho_{1}$ and the norm is the trace norm \cite{helstrom}.  
	
	We are now faced with the problem of choosing $|\psi\rangle$.  Our strategy will be to choose it so that the overlap between $|\psi\rangle$ and either of the states $(U\otimes I)|\psi\rangle$ or $(I\otimes U)|\psi\rangle$ is as small as possible.  This makes the states where one detector has interacted and the state in which none have as distinguishable as possible.  Setting $|\psi\rangle = \sum_{j,k=\pm} c_{jk}|u_{j}\rangle |u_{k}\rangle$, we have 
	\begin{eqnarray}
	\langle\psi |(U\otimes I)|\psi\rangle & = & (|c_{++}|^{2} + |c_{+-}|^{2})e^{i\theta} \nonumber \\
	&& + (|c_{-+}|^{2}+|c_{--}|^{2}) e^{-i\theta} \nonumber \\
	\langle\psi |(I\otimes U)|\psi\rangle & = & (|c_{++}|^{2} + |c_{-+}|^{2})e^{i\theta} \nonumber \\
	&&+ (|c_{+-}|^{2}+|c_{--}|^{2}) e^{-i\theta} .
	\end{eqnarray}
	For a fixed $\theta$, the magnitudes of both of these expressions are minimized when all of the $c_{jk}$ are the same, so that for $|\psi\rangle$ we choose the product state
	\begin{equation}
	|\psi\rangle = \frac{1}{2} (|u_{+}\rangle + |u_{-}\rangle ) (|u_{+}\rangle + |u_{-}\rangle ) .
	\end{equation}
	
	We now have to compute the trace norm of $\Lambda$, which means we have to diagonalize it.  Defining $|v_{\pm}\rangle = (1/\sqrt{2})(|u_{+}\rangle \pm |u_{-}\rangle )$ we can express $\Lambda$ in the basis $\{ |v_{+}\rangle |v_{+}\rangle  ,  |v_{+}\rangle |v_{-}\rangle ,  |v_{-}\rangle |v_{+}\rangle \}$
	\begin{equation}
	\Lambda = \left( \begin{array}{ccc} p_{0}-p_{1}c^{2} & iscp_{1}/2 & iscp_{1}/2 \\ -iscp_{1}/2 & -p_{1}s^{2}/2 & 0 \\ -iscp_{1}/2 & 0 & -p_{1}s^{2}/2 \end{array} \right) ,
	\end{equation}
	where $c=\cos\theta$ and $s=\sin\theta$.  From this we can find the eigenvalues, and from them we find that the trace norm is
	\begin{equation}
	\|\Lambda\|_{1} = \frac{1}{2} \left[ p_{1}^{2}(1+c^{2})^{2} + 4p_{0}^{2} + 4p_{0}p_{1}(1-3c^{2}) \right]^{1/2} + \frac{p_{1}s^{2}}{2} .
	\end{equation}
	For small values of $\theta$, the success probability is most sensitive to $\theta$ at $p_{0}=p_{1}=1/2$.  There we find that $P_{s}\cong (1/2)[1+ (\theta /\sqrt{2})]$.  Once we get away from equality for the probabilities, the leading term in $\theta$ is quadratic rather than linear.  Note that this implies that the measurement gives us the most information in the case in which the classical information is least; if $p_{0}=p_{1}=1/2$, then \emph{a priori} we have no information about which alternative occurred, and this is where the measurement helps the most.
	
	 Note that this means that a given detector array can be flexible.  The physical array can remain the same, but the best measurement and initial stated depend on the desired question being asked.
	
	\section{N detectors}
	\subsection{Finding which detector interacted}
	Now let us look at the case of $N$ detectors, and initially we will assume that only one has registered something, and we want to find out which one.  We are going to assume that the state to which the detectors will be applied is a symmetric state.  A symmetric multi-qubit state is one that is invariant under permutations of the qubits.  Our Hilbert space has a basis consisting of products of $N$ qubit states, where each qubit is in the state $|u_{+}\rangle$ or $|u_{-}\rangle$.  Now let $|k;N\rangle$ be a normalized state of $N$ qubits that is an equal superposition of all the basis elements with $k$ $|u_{+}\rangle$ states.  If our initial state is $|k;N\rangle$, the detector states are given by the application of the operators $F_{n} = I^{\otimes (n-1)} \otimes U\otimes I^{\otimes (N-n)}$ to $|k;N\rangle$.  The inner product between two different detector states is
	\begin{equation}
	\label{innprod1}
	\langle k;N|F_{n}^{\dagger} F_{m}|k;N\rangle = 1 -  \frac{1}{N(N-1)} (2k)(N-k) (1-\cos 2\theta ) .
	\end{equation}
	Note that this does not depend on $m$ and $n$.  As a function of $k$ this is a minimum when $k=N/2$ for $N$ even and $k=(N-1)/2$ (or $(N+1)/2$) for $N$ odd.  This suggests that among the states $|k;N\rangle$, this choice for $k$ is the best, because the resulting detector states are the most distinguishable.  
	
	It is useful to compare this choice of $|\psi\rangle$ to one that is simply a product state,
	\begin{equation}
	|\psi_{sep}\rangle = \left(\frac{1}{2} \right)^{N/2} (|u_{+}\rangle + |u_{-}\rangle )^{\otimes N} .
	\end{equation}
	In that case, we find that
	\begin{equation}
	\langle\psi_{sep}|F_{n}^{\dagger}F_{m}|\psi_{sep}\rangle = 1-\frac{1}{2} (1-\cos 2\theta ) .
	\end{equation}
	Comparing this to Eq.\ (\ref{innprod1}) with $N$ even and $k=N/2$ (see Eq.\ (\ref{innprod2}) below), we see that the inner product is smaller for the entangled state, but this difference goes to zero as $N$ increases.  That means the states resulting from an initial entangled state are more distinguishable, but it also suggests that the advantage of using entangled states is greatest for a small number of detectors.
	
	We are then reduced to the problem of discriminating a set of states any two of which have the same inner product.  Using the pretty-good measurement, this problem has been solved in \cite{englert}.  Adopting the notation in that paper, and assuming for simplicity that $N$ is even, let $|E_{j}\rangle = F_{j}|(N/2);N\rangle$, for $j=1,2,\ldots N$ and $|H\rangle =(1/N)\sum_{j=1}^{N} |E_{j}\rangle$.  Note that for $j\neq k$,
	\begin{equation}
	\label{innprod2}
	\langle E_{j}|E_{k}\rangle = 1-\frac{N}{2(N-1)} ( 1 - \cos 2\theta ) .
	\end{equation}
	Define
	\begin{eqnarray}
	\label{r0}
	r_{0} & = & \langle E_{j}|H\rangle = \langle H|H\rangle = \frac{1}{2}(1+\cos 2\theta ) \nonumber \\
	r_{1} & = & r_{0} - \langle E_{j}|E_{k}\rangle \hspace{5mm} {\rm for}\ j\neq k  \nonumber \\
	& = & \frac{1}{2(N-1)} (1-\cos 2\theta ) ,
	\end{eqnarray} 
	which allows us to write
	\begin{equation} 
	\langle E_{j}|E_{k}\rangle = r_{0}-r_{1} + \delta_{jk} Nr_{1} .
	\end{equation}
	In addition, we will need the orthonormal set of $N$ vectors
	\begin{equation}
	|e_{j}\rangle = \frac{1}{\sqrt{Nr_{1}}} ( |E_{j}\rangle - |H\rangle ) + \frac{1}{\sqrt{Nr_{0}}} |H\rangle .
	\end{equation}
	The results in \cite{englert} yield the POVM operators $\Pi_{j}=|e_{j}\rangle\langle e_{j}|$ and the success probability
	\begin{eqnarray}
	P_{s} & = & \frac{1}{N} (\sqrt{r_{0}} + (N-1)\sqrt{r_{1}})^{2} \nonumber \\
	& = & \frac{1}{N} \left[ 1+\frac{1}{2}(N-2) (1-\cos 2\theta )+ \sqrt{N-1} |\sin 2\theta | \right]   . \nonumber \\
	\end{eqnarray}
	In the large $N$ limit, this goes to $(1/2)(1-\cos 2\theta )$.  If we instead use the separable state, $|\psi_{sep}\rangle$, we find
	\begin{eqnarray}
	r_{0} & = & \frac{1}{N} + \frac{N-1}{2N} (1+\cos 2\theta ) \nonumber \\
	r_{1} & = & \frac{1}{2N}(1-\cos 2\theta ) ,
	\end{eqnarray}
	and the success probability for the separable state is
	\begin{eqnarray}
	P_{s}^{(sep)} & = & \frac{1}{N} \left\{ 1 + \frac{(N-1)(N-2)}{2N}(1-\cos 2\theta ) \right. \nonumber \\
	& = & \left. \frac{N-1}{\sqrt{N}} [ \sin^{2} 2\theta + (1/N)(1-\cos 2\theta )^{2} ]^{1/2} \right\} .
	\end{eqnarray}
	We can compare the results of entangled and separable states by looking first at a small $N$ case, in particular $N=2$.  We find
	\begin{eqnarray}
	P_{s} & = & \frac{1}{2} (1 + |\sin 2\theta | ) \nonumber \\
	P_{s}^{(sep)} & = & \frac{1}{2} \left\{ 1 + \frac{1}{\sqrt{2}} [\sin^{2} 2\theta + (1/2) (1-\cos 2\theta )^{2}]^{1/2} \right\} , \nonumber  \\
	\end{eqnarray}
	(the success probability without the superscript is the one from the entangled state) and we can verify that $P_{s} \geq P_{s}^{(sep)}$.  In the large $N$ limit, the difference between the two goes to zero, in particular,
	\begin{eqnarray}
	P_{s} & \rightarrow & \frac{1}{2}(1-\cos 2\theta ) + \frac{1}{\sqrt{N}} |\sin 2\theta | \nonumber \\
	&&+ \frac{1}{N}  - \frac{1}{N}(1-\cos 2\theta ) \nonumber \\
	P_{s}^{(sep)} & \rightarrow & \frac{1}{2}(1-\cos 2\theta ) + \frac{1}{\sqrt{N}} |\sin 2\theta | \nonumber \\
	&& + \frac{1}{N}  - \frac{3}{2N}(1-\cos 2\theta ) ,
	\end{eqnarray}
	where both expressions contain terms up to order $1/N$.
	
	Returning to the entangled state case, for unambiguous discrimination, we define the vectors
	\begin{equation}
	|\bar{e}_{j}\rangle = |e_{j}\rangle + \frac{t-1}{\sqrt{Nr_{0}}} |H\rangle ,
	\end{equation}
	where $1\geq r_{0}>(1/N)>r_{1}$ and $t=\sqrt{r_{1}/r_{0}}$.  The POVM elements are $\Pi_{j}=|\bar{e}_{j}\rangle\langle\bar{e}_{j}|$, and the failure probability is
	\begin{equation}
	P_{f} = \frac{Nr_{0}-1}{N-1} = 1-\frac{N(1-\cos 2\theta )}{2(N-1)} ,
	\end{equation}
	which goes to $(1/2)(1+\cos 2\theta )$ for $N$ large.  
	
	\subsection{Adding the no-interaction state}
	It is possible to add an additional state in the case where the measurement can make errors.  In particular, we will add the state in which no detectors fire, $|E_{0}\rangle = |(N/2);N\rangle$.  This cannot be done in the case of unambiguous discrimination, because for that to be possible, the states must be linearly independent.  The state $|E_{0}\rangle$ is in the space, $\mathcal{H}_{E}$, spanned by the states $|E_{j}\rangle = F_{j}|(N/2);N\rangle$, $j=1,2,\ldots N$.  A short calculation shows that $\sum_{j=1}^{N} |\langle e_{j}|E_{0}\rangle |^{2} = 1$ and that 
	\begin{equation}
	|E_{0}\rangle = \frac{1+e^{-i\theta}}{2\sqrt{r_{0}}} |\tilde{H}\rangle , 
	\end{equation}
	where $|\tilde{H}\rangle = (1/\sqrt{r_{0}}) |H\rangle$ is a normalized version of $|H\rangle$ (note that here $r_{0}$ and $r_{1}$ are given by Eq.\ (\ref{r0})).  We will assume that the probability of no detectors firing is $p$ and that the probability of each of the detectors firing is $(1-p)/N$.  We will exclude the case of more than one detector firing.
	
	In order to find the POVM and success probability, we will make use of the pretty-good measurement \cite{hausladen}.  This does not necessarily give an optimal measurement to discriminate states, but it does give, as its name implies, a pretty good measurement.  If the states to be discriminated are $|\phi_{j}\rangle$, $j=1,2,\ldots M$, where $|\phi_{j}\rangle$ occurs with probability $p_{j}$, then the POVM elements are
	\begin{equation}
	\Pi_{j}=p_{j} \rho^{-1/2}|\phi_{j}\rangle\langle\phi_{j}|\rho^{-1/2} ,
	\end{equation}
	where $\rho = \sum_{j=1}^{M} p_{j} |\phi_{j}\rangle\langle\phi_{j}|$.  In our case
	\begin{eqnarray}
	\rho & = & p |\tilde{H}\rangle\langle\tilde{H}| + \frac{1-p}{N} \sum_{j=1}^{N} |E_{j}\rangle\langle E_{j}| \nonumber \\
	& = & [p+(1-p)r_{0}] |\tilde{H}\rangle\langle\tilde{H}| + (1-p)r_{1} P ,  \nonumber
	\end{eqnarray}
	where $P$ is the projection onto the subspace in $\mathcal{H}_{E}$ orthogonal to $|\tilde{H}\rangle$.  Defining 
	\begin{eqnarray}
	D_{0} & = & \frac{1}{[p+(1-p)r_{0}]^{1/2}} \nonumber \\
	D_{1} & = & \frac{1}{[(1-p)r_{1}]^{1/2}} ,
	\end{eqnarray}
	we have that 
	\begin{eqnarray}
	\rho^{-1/2} & = & D_{0}  |\tilde{H}\rangle\langle\tilde{H}| + D_{1} P \nonumber \\
	\rho^{-1/2} |\tilde{H}\rangle & = & D_{0} |\tilde{H}\rangle \nonumber \\
	\rho^{-1/2} |E_{j}\rangle & = & \sqrt{r_{0}} (D_{0} - D_{1}) |\tilde{H}\rangle + D_{1}|E_{j}\rangle .
	\end{eqnarray}
	From these, we can construct the POVM 
	\begin{eqnarray}
	\Pi_{0} & = & p \rho^{-1/2}|\tilde{H}\rangle\langle \tilde{H}|\rho^{-1/2} \nonumber \\
	\Pi_{j} & = & \frac{1-p}{N} \rho^{-1/2}|E_{j}\rangle\langle E_{j} |\rho^{-1/2} ,
	\end{eqnarray}
	and the success probability
	\begin{eqnarray}
	P_{s} & = & p\langle E_{0}|\Pi_{0}|E_{0}\rangle + \frac{1-p}{N} \sum_{j=1}^{N} \langle E_{j}|\Pi_{j}|E_{j}\rangle \nonumber \\
	& = & p^{2} D_{0}^{2} + \frac{(1-p)^{2}}{N} (r_{0}D_{0}+(1-r_{0})D_{1})^{2} .
	\end{eqnarray}
	In the large $N$ limit this becomes
	\begin{equation}
	P_{s}\rightarrow \frac{1}{2}(1-p) (1-\cos 2\theta ) + \frac{p^{2}}{p+(1/2)(1-p)(1+\cos 2\theta )} .
	\end{equation}
	
	\section{Conclusion}
	Two key issues in the study of quantum detector networks are the nature of the best measurements to gain information about the question at hand and the nature of the best initial state of the detectors.  In particular, when is an entangled state best?  As noted in the Introduction, in the case of parameter estimation, entanglement did not provide an advantage is estimating parameters associated with individual detectors, but did when estimating parameters associated with several detectors, for example, the sum of parameters for individual detectors.  We studied a different problem, and we have presented a model of a detector network as a set of detectors that either do, or don't interact with the environment. Assuming at most one detector does interact, we wish to determine which detector has interacted.  As noted, there are two aspects to this problem.  The first is choosing the initial state of the detectors.  In the case of two detectors, the problem can be solved completely, and an entangled initial state is optimal.  For more than two detectors, some assumptions are in order, because exact solutions are not known.  We looked at the case of separable and entangled initial states and found that the entangled state gave an advantage, but its advantage decreased as the number of detectors increased.  Since entangled states are harder to produce than separable ones, this suggests that for small numbers of detectors the use of an entangled initial state is worth the cost, but for larger numbers it is not.  The second aspect is choosing the measurement.  We considered global measurements in which all of the detectors are measured at once, and we made use of the pretty-good measurement to find a POVM and the success probability.  We found that the success probability went to a finite limit as the number of detectors becomes large.  If we simply guessed which detector interacted, our probability of success would be $1/N$, which goes to zero in the large $N$ limit, so the measurement is a major improvement over the guessing result.  In the case in which the no-interaction state is included, for $p>1/(N+1)$ the guess probability is just $p$ (just guess the most probable state), and the success probability of the measurement is greater.  If $p< 1/(N+1)$, the guess probability is between $1/N$ and $1/(N+1)$, which goes to zero in the large $N$ limit, while the success probability of the measurement goes to a constant.  
	
	In regard to the role of entanglement, we find that whether it gives an advantage depends on the question being asked and on the circumstances.  In the case of determining whether one of two detectors has interacted or neither has, entanglement does not seem to help, whereas if one is trying to determine which of two detectors has interacted it does.  For more than two detectors, there is an advantage, but it decreases as the number of detectors increases.  This suggests that it would be worthwhile to study other other situations, for example dividing the detectors into sets and trying to determine in which set an interacting detector is located. Does entanglement help here?  This is reserved for future work.
	
	\acknowledgments
	This research was supported by NSF grant FET-2106447.

\end{document}